\documentclass{kapproc}
\normallatexbib

\begin{document}
\input epsf 

\articletitle{On the Berezin Description of K\"ahler Quotients}

\author{Iliana Carrillo-Ibarra}
\affil{Departamento de Matem\'aticas\\
Centro de Investigaci\'on y de Estudios Avanzados del IPN\\
Apdo. Postal 14-740, 07000, M\'exico D.F., M\'exico}
\email{{\tt iliana@math.cinvestav.mx}}

\author{Hugo Garc\'{\i}a-Compe\'an}
\affil{Departamento de F\'{\i}sica\\
Centro de Investigaci\'on y de Estudios Avanzados del IPN\\
Apdo. Postal 14-740, 07000, M\'exico D.F., M\'exico}
\email{{\tt compean@fis.cinvestav.mx}}

\chaptitlerunninghead{Berezin's Quantization of K\"ahler Quotients}

\begin{abstract}
We survey geometric prequantization of finite dimensional affine K\"ahler manifolds.
This prequantization and the Berezin's deformation quantization formulation, as proposed 
by Cahen {\it et al.}, is used to quantize
their corresponding K\"ahler quotients. Equivariant formalism greatly facilitates the description.  

\end{abstract}


\section{Motivation}

Chern-Simons (CS) gauge theory in 2+1 dimensions is a very interesting quantum field theory which has
been very useful to describe diverse sorts of physical and mathematical systems. On the physical side, we
have the fractional statistics of quasi-particles in the fractional quantum Hall effect \cite{wilczek},
Einstein gravity in 2+1 dimensions with nonzero cosmological constant \cite{achu}. On the mathematical side
it is related to beautiful mathematics like knot and link invariants \cite{cs} and to quantum groups
\cite{qg}.
There is also a nice relation with conformal field theory in two dimensions \cite{cs,eli}. From the
quantization of this
theory we also have learned a lot of things like a very non-trivial generalization of the original
Jones representations of the braid group \cite{adw}.

\vskip .5 truecm
%%%%%%%%%%%%%%%%%%%%%%%%%%%%%%%%%%%%%%%%%%%
\noindent
{\it Canonical Quantization}

In 2+1 dimensions CS gauge theory is based in the Lagrangian 

\begin{equation}
L_{CS} = {k \over 4 \pi}\int_M Tr \bigg( A \wedge d A + {2 \over 3} A \wedge A \wedge
A \bigg),
\end{equation}
where $A$ is a Lie algebra ($Lie(G)$) valued gauge connection on the $G$-bundle $E$ over $M$
and where $G$ is a compact and simple finite dimensional Lie group, e.g. $SU(2)$. Thus $A=
\sum_{a=1}^{dim \ G} A^a_I T_a dx^I$ (with $I,J=0,1,2$) where $T_a$ are the generators of $Lie(G)$ and
$k \in {\bf Z}$ is the level of the theory.

In the classical theory the equations of motion are given by the ``flat connection'' condition

\begin{equation} 
F(A) = dA + A \wedge A =0 \Longleftrightarrow F^a_{IJ} =0.
\end{equation}
While the constraints are given by the Gauss law: 

\begin{equation}
{\delta L \over \delta A^a_0} = \varepsilon_{IJ} F_{IJ}^a =0.
\end{equation}
Canonical quantization on $M= \Sigma \times {\bf R}$ consists in the construction of a 
Hilbert space ${\cal H}_{\Sigma}$  associated to the two-dimensional surface $\Sigma$.
The construction of ${\cal H}_{\Sigma}$ is as follows. First, the decomposition of $M$ 
allows to fix the temporal gauge $A_0 = 0$. In this gauge, the Poisson bracket is given by

\begin{equation}
\{ A^a_I(x), \pi^b_J(y) \} = {4 \pi \over k} \varepsilon_{IJ} \delta^{ab} \delta^2(x-y),
\end{equation}
where $\pi^a_I (x) = {\partial L \over \partial \dot{A}^a_I} = A^a_I(x)$. 

The  phase space ${\cal A}$ consist of the solutions of Eq. (2). That means all flat
connections on $\Sigma$. The incorporation of the  constraints (3) leads to the moduli
space of certain families of vector bundles over $\Sigma$.  This can be identified with the 
reduced phase space ${\cal M} = {\cal A}/{\cal G}$. Pick a complex structure $J$ on
$\Sigma$ leads to consider to ${\cal M}$ as a compact K\"ahler manifold ${\cal M}_J$ of
finite dimension. This space is precisely the moduli space of certain family of holomorphic
vector bundles over $\Sigma$. The quantization of the manifold ${\cal M}_J$ leads to the 
quantization of the Chern-Simons theory given by the Lagrangian (1). This can be
constructed with the help of the Quillen determinant line bundle ${\cal L}$.  This is a
line bundle over ${\cal M}_J$ whose first Chern class $c_1({\cal L})$ coincides with the 
simplectic form $\omega_0$ which determines the Poisson brackets $\{A^a_I(x), A^b_J(y)\} =
\omega_0^{-1}(dA,dA).$ For the arbitrary level $k$ the symplectic form $\omega = k \omega_0$ 
is given by the first
Chern class of the $k$-th power of the line bundle ${\cal L}^{\otimes k}$. Finally the
Hilbert space ${\cal H}_{\Sigma}$ is constructed from the space of $L^2$-completion of
holomorphic sections 
$H^0_{L^2} ({\cal M}_J,{\cal L}^{\otimes k})$ of the determinant line bundle ${\cal L}^{\otimes k}$.  

The quantization of the CS theory is given by the geometric quantization of the
physical reduced phase space ${\cal M}_J$. This was given by Axelrod, Della Pietra and
Witten \cite{adw}. 

In the present note we use the Berezin's deformation quantization formalism
\cite{berezin,cahenuno,cahendos,tajan,martin} to quantize the reduced phase space $({\cal M}_J,
\tilde{\omega})$
which can be regarded as a finite dimensional K\"ahler quotient space. This is a preliminary step to 
apply the Berezin's formalism to quantize the more involved infinite dimensional case of Chern-Simons
gauge theory in three dimensions with compact groups. One possible guide to address
the infinite dimensional
case would be the case of the quantization of ${\bf CP}^{\infty}$ \cite{qm}.
This
will be reported in a forthcoming paper \cite{hugo}. Recent interesting applications of Berezin's
quantization are found in \cite{ncsolitons,isidro}.

%%%%%%%%%%%%%%%%%%%%%%%%%%%%%%%%%%%%%%%%%%%%%%%%%%%%%%%%%%%%
\section{Geometric Quantization of K\"ahler Quotients}

\noindent
{\it Up-stairs Geometric Quantization}

We first consider the finite dimensional affine symplectic manifold 
$({\cal A}, \omega)$ with a chosen complex structure $J$ on ${\cal A},$ invariant under affine
translations \cite{adw}. This 
induces a K\"ahler structure on ${\cal A}$. In order to quantize the affine K\"ahler
space ${\cal A}_J$ we consider the prequantum
line bundle ${\cal L}$ over ${\cal A}$ with connection $\nabla$. This connection has curvature
$[\nabla,\nabla] = -i \omega$ 
since symplectic connection $\omega$ is of the $(1,1)$ type and type $(0,2)$ component of the curvature
vanishes.
This induces a holomorphic structure on $({\cal L}, \nabla)$. Consider the $L^2$-completion 
of the subset of holomorphic sections ${\cal H}_{\cal Q}|_J \equiv H^0_{L^2}({\cal A}_J, {\cal L})$ of
the prequantum Hilbert space $H^0_{L^2}({\cal A}, {\cal L})$. This
constitutes the
Hilbert space of the quantization of $({\cal A}_J, \omega)$. The K\"ahler quantization depends on the 
choice of $J$. In the theory of geometric quantization it is impossible to choice in a natural way
a K\"ahler polarization. That is, to choice a complex structure $J$ for which $\omega$ has the properties 
of a K\"ahler form. Thus the K\"ahler polarization is not unique. In the procedure of quantization we 
should make sure that the final result will be independent on this complex structure $J$ and it depends 
only on the underlying symplectic geometry. Thus the idea is to find a canonical identification of
the ${\cal H}_{\cal Q}|_J$ as $J$ varies (for further details see \cite{adw}).

\vskip .5truecm
%%%%%%%%%%%%%%%%%%%%%%%%%%%%%%%%%%%%%%%%
\noindent
{\it Down-stairs Geometric Quantization}

Now we will consider symplectic quotients of finite-dimensional affine symplectic spaces. The idea is
geometric quantize the reduced phase space $({\cal M}_J, \widetilde{\omega})$ where ${\cal M}_J= {\cal
A}_J//{\cal G}$ is the {\it Marsden-Weinstein quotient}.

We start from $({\cal A}_J,\omega)$ with the action of the group ${\cal G}$ acting on ${\cal A}_J$ by
symplectic diffeomorphisms. 

Let ${\bf g}$ be the Lie algebra of ${\cal G}$ and consider the map
$T: {\bf g} \to Vect({\cal A}_J)$. Since ${\cal G}$ preserves the symplectic form, 
the image of $T$ is a subspace of $Vect({\cal A}_J)$ consisting in the symplectic vector fields on ${\cal
A}_J$.
The co-moment map is given by $F: {\bf g} \to Ham({\cal A}_J)$ where  $Ham({\cal A}_J)$ is the space of 
Hamiltonian functions on ${\cal A}_J$. Take a basis of ${\bf g}$ to be
$\{L_a\} $ and we have $T_a = T(L_a)$. $T$ is a Lie algebra representation since $[T_a,T_b] =
f_{ab}^cT_c$ with $a,b,c = 1, \dots , dim{\cal G}$. 

For each $x \in {\cal A}_J$, $F_a(x)$ are the components of a vector in the dual
space ${\bf g}^V$. That means that there is a maping $F: {\cal A}_J \to {\bf g}^V$ called 
the {\it moment map}. $F^{-1}(0)$ is ${\cal G}$-invariant so one can define the symplectic quotient
of ${\cal A}_J$ and ${\cal G}$ as ${\cal M}_J = F^{-1}(0)/{\cal G}$. Thus one have 
$\pi : F^{-1}(0) \to F^{-1}(0)/{\cal G} \equiv {\cal M}_J$
where $x \mapsto \widetilde{x}$. ${\cal M}_J$ also have structure of a symplectic manifold
whose symplectic structure $\widetilde{\omega}$ is given by
$\widetilde{\omega}_{\widetilde{x}}
(\widetilde{V}, \widetilde{W}) = \omega_x(V,W)$ for $\widetilde{V}, \widetilde{W} \in
T_{\widetilde{x}}{\cal A}_J$.

We consider on ${\cal A}_J$ only ${\cal G}$-invariant quantities so that when restricted to $F^{-1}(0)
\subset {\cal A}_J$ they are pushed-down to the corresponding objects in ${\cal M}_J$. 

The prequantum line bundle can be pushed-down as follows. The symplectic action of ${\cal G}$ on
${\cal A}_J$ can be lifted to ${\cal L}$ in such a way that it preserves the connection and
Hermitian structure on ${\cal L}$. One may define the pushdown bundle $\widetilde{\cal L}$ by stating
that its
sections $\Gamma({\cal M}_J, \widetilde{\cal L}) \equiv \Gamma(F^{-1}(0), {\cal L})^{\cal G}$ constitutes 
a ${\cal G}$-invariant subspace of the space of sections $\Gamma({\cal A}_J,{\cal L})$. The connection
also can be pushed-down and it satisfies 
$\widetilde{\nabla}_{\widetilde{V}} \psi = {\nabla}_{{V}} \psi .$ Meanwhile the curvature of the
connection
$\widetilde{\nabla}$ is $-i \widetilde{\omega}$. Thus the prequantization is given by 
$(\widetilde{\cal L}, \widetilde{\nabla}, \langle \cdot | \cdot \rangle_{\widetilde{\cal L}})$. Here 
$\langle \cdot | \cdot \rangle_{\widetilde{\cal L}}$ is the ${\cal G}$-invariant inner product 
$\langle \cdot | \cdot \rangle^{\cal G}_{\cal L}$. 

%%%%%%%%%%%%%%%%%%%%%%%%%%%%%%%%%%%%%%%%%%%%%%%%%%%%%%%%%%%%
%%%%%%%%%%%%%%%%%%%%%%%%%%%%%%%%%%%%%%%%%%%%%%%%%%%%%%%%%%%%
\section{Berezin Quantization of K\"ahler Quotients}

The main goal of this section is to describe the Berezin's quantization of the K\"ahler manifold
$({\cal M}_J,
\widetilde{\omega})$ where ${\cal M}_J$ is the Marsden-Weinstein quotient. That means we find an
associative and
noncommutative family of algebras $(\widetilde{{\cal S}_B}, \widetilde{*_B})$ with
$\widetilde{{\cal
S}_B} \subset C^{\infty}({\cal M}_J)$ which is indexed with a real and positive parameter $\hbar$ which
helps to recover the classic limit when $\hbar \to 0$. Here we set $\hbar =1.$

In order
to do that we follows the same strategy that for the
geometric quantization case of the previous section. We first Berezin quantize $({\cal A}_J, \omega)$,
{\it i.e.} we find an
associative and noncommutative family of algebras $({\cal S}_B, *_{B})$ with ${\cal
S}_{B}
\subset C^{\infty}({\cal A}_J)$ (see \cite{cahenuno,cahendos}). After that we project out all relevant
quantities to be ${\cal
G}$-invariant
{\it i.e.} $\widetilde{{\cal S}_B} \subset C^{\infty}({\cal A}_J)^{\cal G} \equiv 
C^{\infty}(F^{-1}(0)/{\cal
G})$ with $F^{-1}(0) \subset {\cal A}_J$. 

\vskip .5truecm
%%%%%%%%%%%%%%%%%%%%%%%%%%%%%%%%%
\noindent
{\it Up-Stairs Berezin's Quantization}

Let $({\cal L}, \nabla, \langle \cdot | \cdot\rangle_{\cal L})$ be a prequantization of the
affine K\"ahler manifold ${\cal A}_J$. The inner product $\langle \cdot | \cdot\rangle_{\cal L}$ is
compatible with the connection $\nabla$ and it is defined as

\begin{equation}
\langle \chi | \psi \rangle_{\cal L} \equiv \int_{{\cal A}_J}  \langle \chi | \psi \rangle {\omega^n
\over n!},
\end{equation}
for all $\chi,\psi \in H^0_{L^2}({\cal A}_J,{\cal L})$ where
$\langle \chi | \psi \rangle = \exp \big( -\Phi \big) \overline{\chi} \psi$
and 
$\Phi$ is the K\"ahler potential $\Phi(Z,\overline{Z}) = \sum_i Z^i\overline{Z}^i$. The norm of an
element $\psi$ of
$H^0_{L^2}({\cal A}_J,{\cal L})$ is given by $\langle \psi | \psi \rangle_{\cal L} \equiv ||\psi
||^2_{\cal
L}$. As ${\cal A}_J$ is topologically trivial, the line bundle
${\cal L}$ 
can be identified with the {\it trivial} holomorphic line bundle whose holomorphic sections are
holomorphic
functions $\psi$ with Hermitian structure {\it i.e.} $\langle \psi | \psi \rangle = \exp (- \Phi)
|\psi|^2$.
The curvature of the connection compatible with the Hermitian structure is given by $\bar{\partial}
\partial (-\Phi) = \sum_i dZ^i \wedge d\overline{Z}^i = -i \omega$. Of course the existence of a
prequantization 
bundle implies that ${\omega  \over 2 \pi} \in H^2({\cal A}_J, {\bf Z})$.

Take ${\cal Q} \in
{\cal L}_0$ and $\pi[{\cal Q}] = x \in {\cal A}_J$ with local complex coordinates $\{Z^i,
\overline{Z}^i\}$.
Here ${\cal L}_0$ is the line bundle ${\cal L}$ without the zero section.
Now consider $\psi \in H^0_{L^2}({\cal A}_J, {\cal L})$ a holomorphic section such that
${\psi}(x) = {\psi}[\pi({\cal Q})] = L_{\cal Q} ({\psi}) {\cal Q},$
where  $L_{\cal Q} ({\psi})$ is a linear continuous functional of  ${\psi}$.

By the Riesz theorem there is a section $e_{\cal Q} \in H^0_{L^2}({\cal A}_J, {\cal L})$ such
that

\begin{equation}
L_{\cal Q} ({\psi}) = \langle e_{\cal Q}|{\psi} \rangle_{\cal L}
\end{equation}
with ${\cal Q} \in {\cal L}_0$. $e_{\cal Q}$ is known in the literature as
a {\it generalized coherent state}. 

Now consider a bounded operator $\widehat{\cal O} : H^0_{L^2}({\cal A}_J, {\cal L}) \to 
H^0_{L^2}({\cal A}_J, {\cal L}).$ Define the {\it covariant} symbol of this operator as

\begin{equation}
{\cal O}_B(x) = { \langle e_{\cal Q}| \widehat{\cal O}| e_{\cal Q} \rangle_{\cal L} \over
|| e_{\cal Q} ||^2_{\cal L}},
\end{equation}
where ${\cal Q} \in {\cal L}_0$ and $\pi ({\cal Q}) = x$. Define the space of covariant symbols ${\cal
S}_B= \{ {\cal O}_B(x), \ {\rm which \ are \ covariant \ symbols  \
of \ operators} \ \widehat{\cal O} \}$. Each covariant symbol can be analytically continued
to the open dense subset of
${\cal A}_J \times {\cal A}_J$ in such a way
$\langle e_{{\cal Q}} | e_{{\cal Q}'}\rangle_{\cal L} \not= 0$ with $\pi({\cal Q}) =x$ and $\pi({\cal
Q}')
=y$ ( with local coordinates $\{W^i,\overline{W}^i\}$). The obtained symbol is holomorphic in the first
entry
and anti-holomorphic in the second entry. This analytic
continuation is reflected in the covariant symbol in the form

\begin{equation}
{\cal O}_B(Z,\overline{W}) = {\langle e_{{\cal Q}}| \widehat{\cal O}| e_{{\cal Q}'}\rangle_{\cal L} \over
\langle e_{{\cal Q}} | e_{{\cal Q}'}\rangle_{\cal L}}.
\end{equation}

The operator $\widehat{\cal O}$ can be recovered from its symbol in the form

\begin{equation}
\widehat{\cal O} \psi (Z) = \langle e_{{\cal Q}}| \widehat{\cal O}| \psi \rangle_{\cal L} {\cal Q}.
\end{equation}

The consideration of the completeness condition $ {\bf 1} = \int_{{\cal A}_J} | e_{\cal Q} \rangle \langle
e_{\cal
Q}| \exp \big(-\Phi(Z,\overline{Z}) \big) {\omega^n \over n!}(Z,\overline{Z})$ leads to

\begin{equation}
\widehat{\cal O} \psi (Z) = \int_{{\cal A}_J} {\cal O}_B(Z,\overline{W}){\cal B}_{\cal Q}(Z,\overline{W})
\psi
(W) \exp \big( - \Phi(W,\overline{W}) \big)
{\omega^n \over n!}   
(W,\overline{W}) {\cal Q},
\end{equation}
where $\psi(W) = \langle e_{{\cal Q}'} | \psi \rangle_{\cal L},$ $\pi[{\cal Q}] = x,$ ${\cal Q} \in
{\cal L}_0$ and ${\cal B}_{\cal Q}(Z,\overline{W}) \equiv 
\langle e_{{\cal Q}} | e_{{\cal Q}'}\rangle_{\cal L}.$ ${\cal B}_{\cal Q}(Z,\overline{W})$ is the
generalized Bergman kernel.

In order to connect this global description with the (local) standard  Berezin formalism is important to
set
a dense open subset ${\cal U}_J$ of the affine space ${\cal A}_J$. Then there is a holomorphic
section $\psi_0 : {\cal U}_J \to {\cal L}_0$ and a holomorphic function  $\phi: {\cal U}_J \to {\bf C}$
such that $\psi(x) = \phi(x) \psi_0(x)$ with $x \in {\cal U}_J \subset {\cal A}_J$. Define the map 
${\cal U}_J \to H^0_{L^2}({\cal U}_J,{\cal L})$ such that $x \mapsto \phi_x$. Let $\phi_x$ and 
$\phi '$ be two elements
of $H^0_{L^2}({\cal U}_J , {\cal L})$ then 

\begin{equation}
\phi'(x) = \langle \phi '| \phi_x \rangle_{\cal U} = 
\int_{{\cal U}_J} \phi '(y) \overline{\phi}_x(y) |\psi_0|^2(y) \exp \big( - \Phi(y)) {\omega^n
\over n!} (y),
\end{equation}
where  $\langle \cdot| \cdot \rangle_{\cal U}$ is the inner product in $H^0_{L^2}({\cal U}_J , {\cal L})$.

Let $\widehat{\cal O}$ be a bounded operator on $H^0_{L^2}({\cal
A}_J,{\cal L})$. Define the corresponding operator $\widehat{\cal O}_{0}$ acting on $H^0_{L^2}({\cal
U}_J,{\cal L})$ by $\widehat{\cal O} \psi = [\widehat{\cal O}_{0} \phi] \psi_0$ where $\psi \in
H^0_{L^2}({\cal A}_J, {\cal L})$ and $\psi = \phi \psi_0$ on ${\cal U}_J$.

The analytic continuation of the covariant symbol when restricted to ${\cal U}_J \times
{\cal U}_J$ is given by

\begin{equation}
{\cal O}_{B(0)}(Z,\overline{W}) = { \langle \phi_{x} \psi_0| \widehat{\cal O}_0| \phi_y
\psi_0\rangle_{{\cal
U}_J}
\over
\langle \phi_{x} \psi_0| \phi_y \psi_0 \rangle_{{\cal U}_J}} = { \langle \phi_{x}| \widehat{\cal
O}_0| \phi_y \rangle_{{\cal
U}_J} \over
\langle \phi_{x}| \phi_y \rangle_{{\cal U}_J}}.
\end{equation}

The function ${\cal O}_{B(0)}(Z,\overline{Z}) \in C^{\infty}({\cal U}_J)$ is called the
{\it covariant symbol} of the operator $\widehat{O}_0$.
Now if ${\cal O}_{B(0)}(Z,\overline{W})$ and ${\cal O}'_{B(0)}(Z,\overline{W})$ are two covariant symbols
of $\widehat{\cal O}_0$ and $\widehat{{\cal O}'}_0$, respectively, then the covariant symbol of
$\widehat{\cal O}_0\widehat{{\cal O}'}_0$ is given by the {\it Berezin-Wick star product}  
${\cal O}_{B(0)} *_B {\cal O}'_{B(0)}$ given by

$$
({\cal O}_{B(0)}*_B {\cal O}'_{B(0)})(Z,\overline{Z})
$$
$$
=\int_{{\cal U}_J} {\cal O}_{B(0)}(Z,\overline{W}) {\cal O}'_{B(0)}(W,\overline{Z}) {{\cal
B}(Z,\overline{W}){\cal
B}(W,\overline{Z})
\over {\cal B}(Z,\overline{Z})}
\exp \bigg\{- \Phi(W,\overline{W}) \bigg\} {\omega^n \over n!}(W,\overline{W})
$$
\begin{equation}
= \int_{{\cal U}_J} {\cal O}_{B(0)}(Z,\overline{W}) {\cal O}'_{B(0)}(W,\overline{Z}) \exp \bigg\{
{\cal K}(Z,\overline{Z}| W,\overline{W}) \bigg \} {\omega^n \over n!}(W,\overline{W}), 
\end{equation}
where ${\cal K}(Z,\overline{Z} | W,\overline{W}):= \Phi(Z,\overline{W})+ \Phi(W,\overline{Z})-
\Phi(Z,\overline{Z})-
\Phi(W,\overline{W})$ is called the {\it Calabi diastatic function} and ${\cal B}(Z,\overline{Z})$ is the
usual 
Bergman kernel.

Thus we have find the pair $({\cal S}_B, *_B)$ which constitutes the Berezin's quantization of $({\cal
A}_J,\omega)$.

\vskip .5truecm
%%%%%%%%%%%%%%%%%%%%%%%%%%%%%%%%%
\noindent
{\it Down-Stairs Berezin's Quantization}

Finally we are in position to get the desired Berezin quantization of the K\"ahler quotient $({\cal M}_J,
\widetilde{\omega})$. That means to find the family of algebras $(\widetilde{\cal S}_B,
\widetilde{*_B})$. Having the Berezin's quantization $({\cal S}_B,{*}_B)$ of $({\cal A}_J,{\omega})$ and
following the description of the pushed-down prequantization bundle, we restrict ourselves to
$F^{-1}(0) \subset {\cal A}_J$ and consider only ${\cal G}$-invariant quantities.  

Consider $(\widetilde{\cal L}, \widetilde{\nabla}, \langle \cdot | \cdot\rangle_{\widetilde{\cal L}})$
the pushed-down prequantization with $\widetilde{\cal L} = {\cal L}^{\cal G}$ being the ${\cal
G}$-complex line bundle over ${\cal M}_J.$ The inner product $\langle \cdot |
\cdot\rangle_{\widetilde{\cal L}}$ is the ${\cal G}$-invariant product  $\langle \cdot |
\cdot\rangle_{{\cal L}}$ given by 

\begin{equation} 
\langle \widetilde{\chi} | \widetilde{\psi} \rangle_{\widetilde{{\cal L}}} = \langle \chi | \psi
\rangle^{\cal
G}_{\cal L} = \int_{{\cal M}_J} \langle \widetilde{\chi} | \widetilde{\psi} \rangle
{\widetilde{\omega} \over n!} = \langle {\chi} | {\psi} \rangle_{{{\cal
L}}} 
\end{equation} 
for all $\widetilde{\chi},\widetilde{\psi} \in H^0_{L^2}({\cal M}_J,\widetilde{\cal L}) =
H^0_{L^2}(F^{-1}(0),{\cal L})^{\cal G}_J$
where $\langle \widetilde{\chi} | \widetilde{\psi} \rangle = \langle \chi | \psi \rangle$ and
$\widetilde{\omega}$ is preserved by the action of ${\cal G}$, {\it i.e.} $\omega$ is ${\cal
G}$-invariant. The norm of an element
$\widetilde{\psi}$ of $H^0_{L^2}(F^{-1}(0),{\cal L})^{\cal G}_J$ is given by $\langle \widetilde{\psi} |
\widetilde{\psi}
\rangle_{\widetilde{\cal L}} \equiv [||\widetilde{\psi}
||^2]_{\widetilde{\cal L}}$.

Now take $\widetilde{\cal Q} \in
\widetilde{\cal L}_0,$ $\pi[\widetilde{\cal Q}] = \widetilde{x} \in {\cal M}_J$ with local complex
coordinates $\{z^i, \overline{z}^i\}$ and $\pi[\widetilde{\cal Q}'] = \widetilde{y} \in {\cal M}_J$ with
local complex coordinates $\{w^i, \overline{w}^i\}$. Here  $\widetilde{\cal L}_0$ is the line bundle
$\widetilde{\cal L}$ without the zero section. Now
consider $\widetilde{\psi}(\widetilde{x}) = \widetilde{\psi}[\pi(\widetilde{\cal Q})] =
\widetilde{L_{\cal Q}} [\widetilde{\psi}]
\widetilde{\cal Q}$ with  $\widetilde{L_{\cal Q}} [\widetilde{\psi}]$ being a linear functional of
$\widetilde{\psi}$. The group ${\cal G}$ acts on $H^0_{L^2}(F^{-1}(0),{\cal L})_J$ in the
form 

\begin{equation}
(\widetilde{\Gamma} \widetilde{\psi})(\widetilde{x}) \equiv \widetilde{\Gamma} \widetilde{\psi}
(\widetilde{\Gamma}^{-1} \widetilde{x}),
\end{equation}
where $\widetilde{\Gamma} \in {\cal G}$, $\widetilde{x} \in {\cal M}_J$ and $\widetilde{\psi} \in
H^0_{L^2}(F^{-1}(0),{\cal L})^{\cal G}_J.$

Again the Riesz theorem ensures the existence of a section $\widetilde{e_{\cal Q}} \in
H^0_{L^2}(F^{-1}(0),{\cal L})_J^{\cal G}$ such
that $\widetilde{L_{\cal Q}} [\widetilde{\psi}] = \langle \widetilde{e_{\cal Q}}| \widetilde{\psi}
\rangle_{\widetilde{\cal L}}$
with $\widetilde{\cal Q} \in \widetilde{\cal L}_0$. $\widetilde{e_{\cal Q}}$ is the push-down of the
generalized coherent state ${e_{\cal Q}}$. 

Let $\widehat{\cal O}^{\cal G} : H^0_{L^2}(F^{-1}(0), {\cal L})_J^{\cal G} \to 
H^0_{L^2}(F^{-1}(0), {\cal L})_J^{\cal G}$ be a bounded operator. The {\it covariant} symbol of this
operator is defined as

\begin{equation}
{\cal O}_B^{\cal G}(\widetilde{x}) = { \langle \widetilde{e_{\cal Q}}|
\widehat{\cal O}^{\cal G}| \widetilde{e_{\cal Q}} \rangle_{\widetilde{\cal L}} \over [||
\widetilde{e_{\cal Q}} ||^2]_{\widetilde{\cal L}}} \equiv { \langle e_{\cal Q}| \widehat{\cal O}|
e_{\cal Q}
\rangle_{\cal L}^{\cal G} 
\over [|| e_{\cal Q} ||^2]_{\cal L}^{\cal G}},
\end{equation}
where $\widetilde{\cal Q} \in \widetilde{\cal L}_0$ and $\pi (\widetilde{\cal Q}) = \widetilde{x}$.
Now the space of covariant symbols $\widetilde{\cal S_B}$ is defined as the pushing-down of ${\cal
S}_B$,
{\it i.e.} ${\cal S}_B^{\cal G}.$

Similarly to the case of the quantization of $({\cal A}_J,\omega)$, each 
covariant symbol can be analytically continued to the open dense subset of
${\cal M}_J \times {\cal M}_J$ in such a way
$\langle \widetilde{e_{{\cal Q}}}| \widetilde{e_{{\cal Q}'}}\rangle_{\widetilde{\cal L}} \not=
0$ with $\pi(\widetilde{\cal Q}) = \widetilde{x}$ and
$\pi(\widetilde{{\cal
Q}'})
= \widetilde{y}$
which is holomorphic in the first entry and anti-holomorphic in the second entry. This analytic
continuation is written as

\begin{equation}
{\cal O}_B^{\cal G}(z,\overline{w}) = {\langle e_{{\cal Q}}| \widehat{\cal O}| e_{{\cal Q}'}\rangle_{\cal
L}^{\cal
G} \over
\langle e_{{\cal Q}} | e_{{\cal Q}'}\rangle_{\cal L}^{\cal G}}.
\end{equation}

Similar considerations apply to other formulas. But an essential difference with respect to the
quantization of $({\cal A}_J,\omega)$ is that, in the present case, the K\"ahler quotient is
topologically nontrivial and therefore the line bundle $\widetilde{\cal L}$ is {\it non-trivial}. It is
only
locally trivial {\it i.e.} $\widetilde{\cal L}_{(j)} = {\cal W}^{(j)} \times {\bf C}$ for each dense 
open subset ${\cal
W}^{(j)}_J \subset {\cal M}_J$ with $j
=1,2, \dots, N$. Analogous
global formulas found on ${\cal L}$, can be applied only on each local trivialization of
$\widetilde{\cal L}$. Of course,
transition functions on ${\cal W}^{(i)}_J \cap {\cal W}^{(j)}_J$ with $i \not= j$ are very important
and
{\it sections} and other relevant quantities like the {\it Bergman kernel}, {\it K\"ahler potential},
{\it covariant symbols},
etc., transform nicely under the change of the open set (see \cite{qm}). Thus in a particular
trivialization  $\widetilde{\cal L}_{(j)}$ and in the local description,  
the function ${\cal O}^{(j)}_{B(0)}(z,\overline{z}) \in C^{\infty}({\cal W}^{(j)}_J)$ is called the
{\it covariant symbol} of the operator $\widehat{O}^{(j)}_0$.
Now if ${\cal O}^{(j)}_{B(0)}(z,\overline{z})$ and ${\cal O}'^{(j)}_{B(0)}(z,\overline{z})$ are two
covariant
symbols
of $\widehat{\cal O}^{(j)}_0$ and $\widehat{\cal O}'^{(j)}_0$, respectively, then the covariant
symbol of
$\widehat{{\cal O}}^{(j)}_0\widehat{\cal O}'^{(j)}_0$ is given by the {\it Berezin-Wick star
product}  
${\cal O}^{(j)}_{B(0)} \widetilde{*_B} {\cal O}'^{(j)}_{B(0)}$

$$
({\cal O}^{(j)}_{B(0)} \widetilde{*_B} {\cal O}'^{(j)}_{B(0)})(z,\overline{z})
$$
$$
=\int_{{\cal W}^{(j)}_J} {\cal O}^{(j)}_{B(0)}(z,\overline{w}) {\cal O}'^{(j)}_{B(0)}(w,\overline{z})
{{\cal
B}^{(j)}(z,\overline{w}){\cal B}^{(j)}(w,\overline{z}) \over {\cal B}^{(j)}(z,\overline{z})}
\exp \bigg\{- \Phi^{(j)}(w,\overline{w}) \bigg\} {\widetilde{\omega} \over n!}(w,\overline{w})
$$
\begin{equation}
= \int_{{\cal W}^{(j)}_J} {\cal O}^{(j)}_{B(0)}(z,\overline{w}) {\cal O}'^{(j)}_{B(0)}(w,\overline{z})
\exp \bigg\{
{\cal K}^{(j)}(z,\overline{z}|w,\overline{w}) \bigg \} {\widetilde{\omega} \over n!} (w,\overline{w}) 
\end{equation}
where ${\cal K}^{(j)}(z,\overline{z}|w,\overline{w}):= \Phi^{(j)}(z,\overline{w})+
\Phi^{(j)}(w,\overline{z})-
\Phi^{(j)}(z,\overline{z})- \Phi^{(j)}(w,\overline{w})$ is called the {\it Calabi diastatic
function} on ${\cal W}^{(j)}_J$. This construction is valid for all local prequantization
$(\widetilde{\cal L}_{(j)}, \widetilde{\nabla}^{(j)}, \langle \cdot | \cdot\rangle_{\widetilde{\cal
L}_{(j)}})$. Finally, this structure leads to the pair $(\widetilde{{\cal S}_B}, \widetilde{*_B})$ which
constitutes the Berezin
quantization of $({\cal M}_J,\widetilde{\omega})$. 

%%%%%%%%%%%%%%%%%%%%%%%%%%%%%%%%%%%%%%%%%%%%%%%%%%%%%%%%%%%%%%%%%%
%%%%%%%%%%%%%%%%%%%%%%%%%%%%%%%%%%%%%%%%%%%%%%%%%%%%%%%%%%%%%%%%%%
\begin{acknowledgments}
H. G.-C. wish to thank the organizers of the First Mexican Meeting on Mathematical and
Experimental Physics, for invitation. H. G.-C. wish also to thank M. Przanowski and F.
Turrubiates for very useful discussions. I. C.-I. is supported by a CONACyT graduate fellowship.
This work was partially supported by the CONACyT grant No. 33951E. 
\end{acknowledgments}

\end{document}